\documentstyle[aas2pp4]{article}

\slugcomment{Submitted ApJ (letters)}

\lefthead{Wampler}
\righthead{Quasar lenses}

\newcommand {\apgt} {\ {\raise-.5ex\hbox{$\buildrel>\over\sim$}}\, }
\newcommand {\aplt} {\ {\raise-.5ex\hbox{$\buildrel<\over\sim$}}\, }
\makeatletter
\def\section{\@startsection {section}{1}{\z@}{3.5ex plus 1ex minus
    .2ex}{2.3ex plus .2ex}{\large\bf}}
\def\subsection{\@startsection{subsection}{2}{\z@}{3.25ex plus 1ex minus
   .2ex}{1.5ex plus .2ex}{\normalsize\bf}}
\def\subsubsection{\@startsection{subsubsection}{3}{\z@}{3.25ex plus
1ex minus .2ex}{1.5ex plus .2ex}{\normalsize\bf}}
\makeatother

\begin{document}

\title{Do Quasars Lens Quasars?}

\author{E. J. Wampler}
\affil{National Astronomical Observatory Japan, 2-21-2 Osawa, Mitaka, Tokyo 180\\
Japan}

\centerline{accepted, ApJ (Letters)}

\begin{abstract}
If the unexpectedly high frequency of quasar pairs with very different
component redshifts is due to the lensing of a population of background
quasars by the foreground quasar, typical lens masses must be 
$\sim10^{12}\cal M_{\sun}$ and the sum of all such quasar lenses would 
have to contain $\sim0.005$ times the closure density of the Universe. It 
then seems plausible that a very high fraction of all $\sim10^{12}\cal M_{\sun}$ 
gravitational lenses with redshifts $z\sim1$ contain quasars. Here I propose 
that these systems have evolved to form the present population of massive 
galaxies with M$_{\rm B}\leq-22$ and $\cal M$$ >5\times10^{11}\cal M_{\sun}$.
\end{abstract}


\section{Introduction}
Burbidge et al., (1996a) have argued that the surface 
density of random quasars on the plane of the sky is insufficient to account for 
the four known examples of close quasar pairs that have very different component
redshifts. They claim that either these quasar pairs are physically associated
and the discordant redshifts result from very large non-cosmological components,
or that the low redshift quasar marks a gravitational lens that magnifies the 
image of the much higher redshift (background) quasar. I examine the second 
possibility, and develop some of the consequences of lensing by quasars. If these 
four quasars are gravitational lenses, masses $\sim10^{12}$\,$\cal M_{\sun}$ 
are needed to give the image splitting (\cite{is}), a large fraction of all known 
lenses with similar masses contain quasars, and these lenses can plausibly be 
identified as the progenitors of the present day giant galaxies, which contain 
a significant fraction of the total baryonic mass of the Universe.

While other models have also been proposed (\cite{wxp}) in this letter I consider 
lensing by massive galaxies. It will then be seen that very close to the 
entire population of such galaxies is needed to produce the observed statistics. 
These lenses are not soft lenses, like galaxy clusters, as the component 
separations are \aplt5 arcsec. In the following discussion I will consider 
only galaxy-mass gravitational lenses at cosmological distances which magnify 
images of background quasars. I'll use the term ``quasar lens'' to indicate that 
a foreground quasar or a bright Seyfert galaxy is a part of the lens system. By 
``galaxy lens'' I mean a gravitational lens with similar lensing properties but 
without a visible foreground quasar.

\section{Data and observational constraints}
\subsection{Probability of random quasar pairs}
One quasar lens (\cite{jetal}) was discovered accidentally when the telescope 
was inadvertently mispointed during a follow-up observation of its previously 
observed radio companion. A second was identified by \cite{mb} during an HST 
observation of a violently variable quasar, and two more were discovered
by \cite{js} during a survey of bright quasars for gravitational lenses. 

Quasars are rare objects and finding close pairs of unrelated quasars is 
unexpected. Burbidge et al., (1996a) should be consulted for a description of 
the pairs and a detailed discussion of the individual probabilities for each 
pair; but it is easy to see that so many accidental pairs of quasar images are 
unlikely. The number of accidental pairs with separations less than $\theta$ 
arcsec is:
\[ n = 2.42\,\times\,10^{-7}\,\sigma\,\theta^2\,(\sigma\,N)\]
where $n$ is the number of pairs, $N$ is the number of square degrees searched, 
and $\sigma$ is the surface density of quasars on the plane of the sky brighter 
than a given magnitude per square degree. According to Zitelli et al., 
(1992) the surface density of quasars with green (Kron) magnitude $J<20.85$ 
is $\sigma=35\pm7$. Together, the 
two celestial hemispheres lying outside the zone of heavy Galactic absorption 
contain $\sim2\,\times\,10^4$ square degrees. Setting $\theta=5$\,arcsec, the 
number of accidental pairs expected among the entire population of $J<20.85$ 
observable quasars is $n\approx150$. The total number of known quasars brighter 
than 21 mag is about $10^{-2}$ of the total quasar population. Perhaps 1/3 of 
these have been closely examined for close companions. With only $10^{-2.5}$ of 
the quasar population searched for close pairs, the expected number of discovered 
pairs is only 0.5 and it is unlikely that four accidental pairs would already have 
been found if quasars are randomly distributed on the sky.

\subsection{Quasar gravitational lensing}
Although chance alignment of unrelated quasars is unlikely to produce the 4
quasar pairs with greatly discordant redshifts, lensing of the high-redshift 
sky might magnify faint members of the background quasar population enough 
to result in many more pairings (\cite{bhs}). 

For a lens to enhance the apparent density of background quasars, the number-
magnitude count of faint quasars must be sufficiently steep that the increase 
in the numbers of sources caused by the magnification of the lens offsets the 
reduction in area of the background sky caused by the magnification. Following 
\cite{na}, the ratio, $q(M,J_0)$, of the density of objects seen behind 
the lens to the density of unlensed objects is:
\[ q(M,J_0)=\frac{1}{M} \frac{N[<(J_0\,+\,2.5\,log\,M)]}{N(<J_0)}~~.\]
Here N is the number of quasars brighter than magnitude $J_0$ and $M$ is the 
magnification of the lens. The $1/M$ term is the reduction of sky area caused 
by the magnification while the second term is the fractional increase of source 
counts caused by the magnification. The observed brightness of the five high 
redshift companions (one candidate lens has two companions) to the 
four candidate lenses are: 14-19(var), (18.2, 21.2), 17.9, and 18.8. If the 
four discordant-redshift quasar groups are to be explained by lensing then 
$q(M,J_0)\sim8$. Using the quasar counts published by \cite{bo}, \cite{na} 
showed that $q(M,J_0)<3$ if $J_0\apgt18.5$. In this case the lens hypothesis 
would seem to be in difficulty. However, more recently, \cite{bo2} and 
\cite{zi} have found that the faint quasars are more numerous than originally 
estimated by \cite{bo}. This increases the slope of the faint end of the 
quasar log$N$--log$J$ relationship from about 0.3 to 0.4--0.45. With this 
higher slope, $q(10,18.5)=8$ and $q(10,20.5)=1.5$. If the slope were as high 
as .55, the corresponding enhancement factors would be 12 and 2.6. Thus, the 
expected surface density enhancement strongly depends on the surface density 
of the faint quasars near the limits of the present surveys. There seems to 
be sufficient uncertainty in these numbers for the quasar lens hypothesis to 
remain viable.

Despite the uncertainties in the numbers of faint background quasars and
the magnification to be expected, two very general conclusions are valid 
(\cite{pg};\ \cite{js2}): First, the {\em separation} of the quasar pairs 
is a crude measure of the lens mass and is independent of the details of the 
geometry of the Universe. The four discordant-redshift pairs discussed here 
require lens masses that range from $\sim5\times10^{11}\,\cal M_{\sun}$ to 
$\sim5\times10^{12}\,\cal M_{\sun}$. Second, the {\em frequency of occurrence} 
of noticeable lensing is independent of the mass of the individual lenses and 
measures the cosmic density of matter in the direction of the background quasar. 
A gravitational lens focuses light rays, 
and the amount of lens matter along the light-path must ``close'' the Universe 
in that direction. Therefore, the frequency of occurrence of lensing regions 
roughly measures the cosmic density of the lenses in units of the density 
needed to close the entire Universe. In our case the lens occurrence frequency 
is 0.2\%--1\%, which correspond to a similar percentage of the closure density 
of the Universe.

Transparent gravitational lenses produce an odd number of images (\cite{bb}). 
In the magnitude and redshift range that concerns us, the occurrence of 1, 3 
and 5 images is roughly in the ratio 1:1:0.5 (\cite{wn}). One of the 4 
discordant-pair quasars has 2 images displaced from the foreground quasar and 
may be either a 3 or a 5-image lens. The other three candidates could either be 
one or three image lenses. \cite{py} give a simplified description of the geometry 
and amplification expected from spherically symmetric lenses which can produce one 
or three images. In the case of quasar lenses, images close to the lens axis may 
be hard to distinguish from the image of the foreground quasar itself. However,
for Q\,1548+114(A,B), high S/N spectra are available (\cite{sr}), and Iovino and
Shaver (1986) estimate that any secondary images blended with the image of the 
foreground quasar is less than 3\% of the brightness of the high-redshift quasar. 
If Q\,1548+114(A,B) has 3 or more images, the unseen images have little light 
compared to the single observed image of the background quasar. Thus, Q\,1548+114 
might be an example of a gravitational lens producing only a single amplified 
image. 

\subsection{Galaxy gravitational lenses}
There are about 30 candidates for galaxy lenses (\cite{kk}). These lenses are 
presumably all massive individual galaxies or very compact groups of lower mass 
galaxies. But in some cases the lensed images show slightly different spectra 
and/or have slightly different redshifts (\cite{ps}). This subset might consist 
of physical pairs of distinct quasars. Perhaps no more than 25 of the 30 galaxy 
lens candidates are actual gravitational lenses. When comparing the frequency of 
galaxy and quasar lenses, the 25 galaxy lenses should be augmented by a further 
$\sim15$ to account for unrecognized single image lenses, since galaxy lenses 
that produce only single images would not be recognized as gravitational lenses 
while quasar lenses, even with only a single magnified image, would still be 
remarkable. Thus, the relative frequency of quasar to
galaxy lenses is about 1:10. But if quasar light is beamed, as \cite{ba} and
\cite{at}, have suggested, we may see $\aplt\,1/4$ of all the quasars that are 
actually among the gravitational lenses. Therefore, if beaming were important, the 
ratio of quasar lenses to galaxy lenses would be approximately consistent with the 
hypothesis that {\em nearly all} galaxy-mass ($\cal M$$=10^{12}\,\cal M_{\sun}$) 
lenses contain quasars, either beamed towards us (quasar lenses), or beamed away 
from us (galaxy lenses). Even without a correction for unseen quasar lenses, 
1/10 of all $\cal M$$\,\sim\,10^{12}\,\cal M_{\sun}$ lens candidates represents 
a very high frequency of occurrence for quasar lenses.

While total counts are low and selection biases affect quasar and galaxy 
lenses differently, a comparison of quasar lenses with galaxy lenses shows that 
both groups are similar in the separation of the image(s) from the lens axis, 
and in the redshift of the lens, and in the multiplicity of lensed images.
Both the quasar lens candidates and galaxy lens candidates come from similar 
sized parent populations. For both groups about half the candidates have been
discovered accidentally, while the others were uncovered during systematic 
searches. In both groups all the known lenses candidates have been found as 
a result of their association with their 
high redshift companion image(s). It should actually be somewhat harder to 
discover a quasar lens than a galaxy lens since for galaxy lenses all the 
amplified images have the same color and the image of the lens galaxy itself 
is faint and does not contaminate the search field with its light. 

\subsection{Giant galaxies as candidate lenses}
Since the early days of quasar observations (\cite{kr}) it has been known that
quasar images are non-stellar. And spectra of the quasar fuzz shows evidence
for stellar absorption lines (\cite{bor}). Despite the ground-based detection 
of stellar light around quasars, the first imaging observations of quasars with 
the Hubble Space Telescope (HST), seemed to indicate that most quasars were only 
associated with very sub-luminous galaxies \cite{bah}. The apparent discrepancy 
between the HST images and the ground-based images seems now to be understood as 
a result of saturation of the HST quasar image, coupled with uncertainties in 
the HST point-spread-function and the fact that the fuzz light frequently has a 
very smooth spatial distribution (\cite{mr}; \cite{ne}). More recent HST 
observations seem to have overcome most of these difficulties (\cite{hm}) and 
in fact the fuzz brightness around luminous ($J>-24$) quasars is consistent with 
that expected from giant host galaxies (\cite{mr}). Usually the fuzz seems to be 
approximately centered on the quasar, although, as in the case of 3C\,48, 
there are exceptions (\cite{ho}; \cite{hu}).

The cores of present day elliptical galaxies have mass-to-light ratios of about 
8 (\cite{jk}). Thus galaxies with $\cal M$$\,>5\,\times\,10^{11}\,\cal M_{\sun}$ 
have absolute green magnitudes M$_{\rm B}<-22$ (\cite{jk}). The masses of giant 
spiral galaxies are $\sim1/2$ that of ellipticals with the same luminosity. The 
observed image splitting by both quasar and galaxy gravitational lens require 
masses similar to those of nearby giant galaxies (both elliptical and spiral) 
with M$_{\rm B}\leq-22$. Furthermore, the local space density of galaxies with 
M$_{\rm B}<-22$ is $2\times10^{-4}$\,Mpc$^{-3}$(H$_{\rm o}=
50$\,km\,s$^{-1}$\,Mpc$^{-1}$) (\cite{sc}), about 0.005 the closure density of 
the Universe. This also is similar to the density that is needed to produce the 
observed lensing frequency (\cite{bhs};\ \cite{js2}). But a comparison of the local 
space density of bright (M$_{\rm B}<-23$) quasars (\cite{sg}), to that of massive, 
bright (M$_{\rm B}<-22$) galaxies (\cite{sc}) gives only:
\[ \frac{\Phi_{\rm o}(QSO)}{\Phi_{\rm o}(GAL)}\aplt2\times10^{-4}.\]
If, at $z=1$, one tenth of all lenses with masses of $\sim10^{12}$\,$\cal M_{\sun}$
contain visible quasars, the evolution rate between $z=0$ and $z=1$ would have 
to be a factor of $\sim500$. This is in approximate agreement with the observed 
evolution of bright quasars between $z=0$ and $z=1$ (\cite{bo}). Because the
orientations of possible quasar beams would be independent of redshift, any 
unrecognized quasars among the galaxy lenses do not change the required evolution 
rate.

\section{Discussion}
An important advance could be made if the character of the four quasar lens 
candidates were settled. Burbidge et al., (1996) and Wu et al. (1996) have 
examined some of the problems in constructing realistic lens models. They have 
found that it is difficult to produce the required enhancement of background 
quasars without invoking either high density models for the Universe or very 
high galaxy masses. Uncertainties, both in the counts of faint high redshift 
quasars and in the surface density of matter near $z\sim 1$ galaxies, make the 
construction of definitive models unreliable. As already noted, the the simplest 
model -- that the lenses are condensed, high-mass galaxies -- has been accepted 
for this letter. Future data may show that the slope of the faint end of the 
quasar number count is too shallow or that realistic lens mass distributions 
cannot give the needed magnification of the background quasar population. Then
this model would have to be modified or abandoned. But if high resolution imaging, 
together with improved statistics and models, shows that the quasar lens candidates 
are actually lenses, then the ratio of quasar lenses to galaxy lenses will require 
a substantial fraction of all condensed $\sim10^{12}$\,$\cal M_{\sun}$ mass 
concentrations between $z\sim0.5$ and $z\sim1.5$ to contain quasars. As the number 
density of such mass concentrations in comoving coordinates is similar to that of 
present day giant galaxies, it is natural to identify these gravitational lenses 
with the giant galaxies. 

\cite{sf} have argued that elliptical galaxies have little dark matter content, 
at least out to their effective radius. Because it is this inner part of the 
mass distribution that dominates the lens magnification (\cite{wn}), the quasar 
and galaxy lenses should then contain a large fraction of the baryonic matter 
in the Universe. The total baryonic mass density of the Universe is thought to 
lie between 1.3\% and 3\% of the critical density (\cite{st}; \cite{cwc}), 
little higher than that already contained in the giant galaxies. If the quasar 
lenses are not earlier versions of the nearby massive galaxies, then a large 
amount of gravitating material has disappeared with the quasars between $z=1$ 
and the present.

Because the central black hole that is thought to power a quasar is believed 
to have only $\sim10^{-3}$ the mass of a giant galaxy, even low mass galaxies 
might contain quasars. The information needed to characterize the quasar host 
population is a knowledge the host masses. If the quasar lens candidates 
actually turn out to be gravitational lenses, this additional fact supplies the 
needed information. It can then be argued that it would be possible to group the 
quasar lenses and the galaxy lenses into a single population. And this unified 
population of ``quasar lenses'' has evolved to the present set of giant galaxies.
Consistent with the ideas discussed here, McLeod \& Rieke (1995) have shown that 
the fuzz around luminous quasars is as bright as giant early-type galaxies.

The redshifts of the proposed quasar lenses, $0.5\aplt z \aplt1.5$, 
corresponds to a time of strong star-forming activity in the giant 
galaxies (\cite{fs}). The observed high frequency of quasar lenses among the
lens population indicates that the quasar phenomenon is connected to the 
galaxy formation process. At high redshift quasars are often found in rich 
galaxy clusters, while at low redshift they are only found in the field galaxies 
or in poor clusters. This is a natural consequence of a delayed accumulation of 
material in low density regions of the Universe. This past gathering together 
of baryonic mass concentrations $\sim10^{12}\,\cal M_{\sun}$ seems to be both 
a necessary and a sufficient requirement for the presence of a quasar. Smaller 
mass galaxies don't host quasars and high mass galaxies at $z\sim 1$ nearly 
always contain one. 

Finally, if giant galaxies did have a high probability of hosting quasars, the 
quasar lifetimes must have been a substantial fraction of the time it took for 
the Universe to evolve from $z=1.5$ to $z=0.5$; about 5$\times10^9$ years. 

\vspace*{.5cm}

\acknowledgments 
Jack Baldwin, Jean Surdej and an anonymous referee offered helpful criticisms. 
This paper is dedicated to the memories of Bill Burke and Jerry Kristian.

\end{document}